\documentclass[twocolumn]{revtex4}

\usepackage[english]{babel}
\usepackage[letterpaper,top=2cm,bottom=2cm,left=3cm,right=3cm,marginparwidth=1.75cm]{geometry}

\usepackage{amsmath}
\usepackage{graphicx}
\usepackage{subfigure}
\usepackage[colorlinks=true, allcolors=blue]{hyperref}
\usepackage{xcolor}

\def\frep{F_\text{rep}}
\def\hfrep{\hat F_\text{rep}}

\begin{document}

\title{Dynamics of Reversible Plasticity in an Amorphous Solid}
\author{Zhicheng Wang}
\author{Nathan C.\ Keim}
\affiliation{Department of Physics, The Pennsylvania State University, University Park, PA 16802, USA}

\begin{abstract}
Local rearrangements are the elements of plastic deformation in an amorphous solid. In oscillatory shear, they can switch reversibly between two distinct configurations. While these repeating relaxations are typically considered in the limit of slow driving, their dynamics is less well understood. We perform experiments on a colloidal amorphous solid at an oil-water interface. The rearrangement timescales we observe span at least 1 decade, with no apparent upper bound. As frequency is increased, individual rearrangements appear faster and more hysteretic, but may disappear entirely above a crossover frequency---suggesting that in practical experiments, the slowest rearrangements may be latent. We show how to find the effective potential energy that reproduces a particle's frequency-dependent motion. In rare cases, this potential energy has only one minimum. Our results have implications for the energy landscapes and rheology of amorphous or glassy solids, for sound propagation in nonlinear media, and for mechanical memory and history-dependence.
\end{abstract}

\maketitle

\section{Introduction}
The apparent properties of matter often depend on the timescales of the environment or the observer. Silicone ``silly putty'' flows like a viscous liquid when left overnight, but bounces elastically off the floor when dropped. These behaviors arise from the timescales of microscopic structural relaxations in the material, which may be much slower or faster than a given forcing. 
In the laboratory, these relaxation timescales mean that the response to oscillatory shear depends on frequency. In some important cases, such as dilute polymer solutions, the timescales can be predicted from first principles~\cite{rubinstein}.

However, when matter is disordered and crowded, and each relaxation involves structural changes among many elements, the microscopic dynamics involved in deformations are less straightforward. Amorphous solids in which particles lack long-range order---including molecular, colloidal, and polymer glasses, granular packings, and many foams---present such a challenge. 
In static experiments in which these materials age at rest, the typical waiting time between thermally-activated rearrangements grows as the material ages, so that the cumulative relaxation is logarithmic in time and might be limited only by the observer's patience.
Similar aging may also emerge over many cycles of slow loading~\cite{Bandi2018,Shohat2025}.
Nonetheless, these striking long-term results may not be relevant for the dynamics within one cycle, or in steady flow. For example, the widely-used Shear Transformation Zone (STZ) model of amorphous solids typically assumes a single timescale for microscopic relaxations, yet it successfully reproduces experimental observations of viscosity that decreases with strain rate~\cite{Falk1998, Falk2011}. 
Many simulations of deformed amorphous solids attempt to sidestep dynamics entirely, using an athermal, quasistatic approach in which the system comes to static equilibrium after each small strain step.

\begin{figure*}[t]
    \centering
    \includegraphics[width = \textwidth]{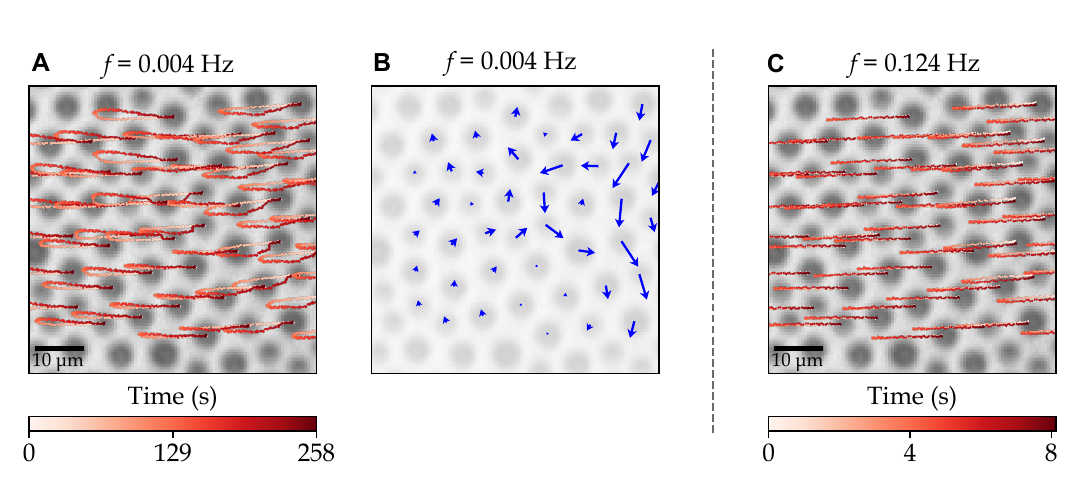}
    \caption{At the microscopic scale, the response of an amorphous solid to oscillatory shear can be plastic or elastic, depending on frequency. 
    \textbf{(A)} Select particle trajectories over a 258-second shear cycle (with maximum shear strain = 0.06; see Fig.~\ref{fig:theory} for waveform) overlaid on experimental micrograph of part of the monolayer. Background image is from the first video frame. Both rearranging (looped trajectories) and non-rearranging (linear trajectories) can be seen. 
    All particles return to the original position after strain returns to 0. Color indicates passage of time. 
    \textbf{(B)} Particle displacements during first half of the cycle in (A) with global affine motion subtracted. Here the displacement vectors shown are 2$\times$ their actual sizes for emphasis purposes. Background image faded for clarity.
    \textbf{(C)} Same particle trajectories over an 8.06-second cycle. In this area, rearranging trajectories are replaced by non-rearranging trajectories. %
    }
    \label{fig:traj}
\end{figure*}

In this paper, we experimentally study a 2D amorphous solid made of repulsive particles in fluid, during cyclic shear over 1.5 decades in frequency. %
We observe hundreds of plastic rearrangements with a broad range of timescales, which repeat with each cycle of shear. %
Within a rearrangement, we treat the coordinated, overdamped motions of particles as one-dimensional. This reveals an effective potential energy that accounts for an individual rearrangement's observed motion at all frequencies. The typical potential is quartic with two wells, but some have only one. %
As in silly putty, dynamics has profound effects: raising the frequency initially accentuates rearrangements' hysteresis, but ultimately renders them inert (Fig.~\ref{fig:traj}). Conversely, lowering the frequency can reveal slow relaxations that were latent. A quasi-static regime with negligible dynamics requires surprisingly low frequencies, beyond the reach of our experiments. Our results suggest that dynamics complicates experiments with amorphous solids, but may lead to rich and largely overlooked possibilities for fundamental studies of glassy matter, and for material memory and computation~\cite{Lindeman2023, Jules2023, Gutierrezprieto2025}. 

\section{Methods}

We performed experiments on a colloid monolayer suspended at a decane-water interface, using an interfacial shear rheometer~\cite{Brooks1999, Qiao2021}. The colloidal monolayer consists of bidisperse polystyrene particles (Invitrogen sulfate latex microspheres, diameters $d_1$ = 4.1~$\mu$m, $d_2$ = 5.4~$\mu$m) at an approximate 50:50 ratio. The combined number density is $\sim$10,200~$\mathrm{mm}^{-2}$, with a surface fraction of  $\sim$18\%. The particles are confined in a 17~mm by 2.4~mm channel with open ends, within a 100~mm diameter glass dish. A magnetized steel needle (spring steel, 0.23~mm diameter, 30~mm long) is magnetically trapped at the middle of the channel, and is moved lengthwise by a computer-controlled magnetic field, so that it applies area-preserving shear to the particle monolayer. The magnetic trap is configured to be relatively stiff, so that the material rheology and needle inertia may be neglected, and the rheometer is effectively strain-controlled: We find that the needle's displacement amplitude decreases by just 3.5\% as the frequency is increased by a factor of 32.

Before a set of experiments we reset the system with a large oscillatory shear cycle ($\gamma_{\text{reset}}=1$) followed by an asymmetric oscillatory ringdown protocol that effectively removes the memory of previous trials~\cite{Keim2022}. %
We then apply oscillatory shear between $\gamma_{\min}=0$ and $\gamma_{\max}=0.06$, at 6 frequencies spaced geometrically from $f = 0.004$~Hz to $f = 0.124$~Hz (period $T = 258$~s to $8.06$~s). Three cycles of oscillatory shear are applied at each frequency, and a waiting period of $t_\text{w}=5$~s follows each cycle at $\gamma_{\min}=0$ both to observe extended relaxation events, and to ensure that the next cycle begins at exactly $\gamma = 0$. We do not include $t_\text{w}$ when stating frequency and period.

A microscope and camera %
record a 3.1~mm by 1.1~mm region (0.78 $\mu$m/px) with $\sim$35,000 particles. The particle trajectories are obtained using trackpy~\cite{trackpy}. We use only particles that are tracked continuously in all 4,977 video frames.

\section{Results}

In Fig.~\ref{fig:traj}(A) we show a small region of the material, over which we plot particle trajectories during oscillatory shear at $0.004$~Hz. Each particle begins and ends the cycle in the same position, but many of the trajectories enclose significant area. These loops belong to localized groups of particles that rearrange during shear. In Fig.~\ref{fig:traj}(B) we isolate the rearranging motions by plotting displacement vectors in the first half of the cycle, after subtracting the prescribed global shear transformation. %
These rearranging regions resemble the structural relaxations observed in many other studies~\cite{Lundberg2008}, and they are known to be associated with local features of the material structure and properties~\cite{Manning2011, Richard2020}, so that in our experiments they recur at the same sites in each repeated cycle of shear. 

\subsection{Overview of kinematics}
To quantify these rearrangements, we use the $D^2_{\min}$ measure of squared non-affine displacement, as defined by Falk and Langer (1998)~\cite{Falk1998,Keim2013,Keim2020,philatracks,supp}. The displacement is measured from the particle positions at time $t$ compared with the reference frame at time $t_0$ where $\gamma=0$. Displacements are rescaled by the typical interparticle distance $a_0\approx 10.16$~$\mu$m measured from centroid to centroid. To obtain the non-affine component, we chose the closest neighbors of each particle (within $2.5 a_0$), and find the best-fit affine transformation that maps the position vectors from time $t_0$ to time $t$. $D^2_{\min}$ is the mean of the squared residuals after fitting.  %
We define $D_{\min} \equiv \sqrt{D^2_{\min}}$ in this work. As in previous work on similar model systems (e.g. Ref.~\cite{Keim2020}), we use
a threshold equivalent to $D_{\min} = 0.12$ to identify particles involved in rearrangements.
When we shear the experimental sample between $\gamma_{\min}=0$ and $\gamma_{\max}=0.06$, these particles constitute roughly 10\% of the total population and are typically found in clusters of roughly 20 particles or more. 
Their spatial distribution is uniform across the imaged section, suggesting a relatively homogeneous bulk deformation, which we verify in the displacement profile across the channel~\cite{supp}. %

\begin{figure}
    \centering
    \includegraphics[width = 0.48\textwidth]{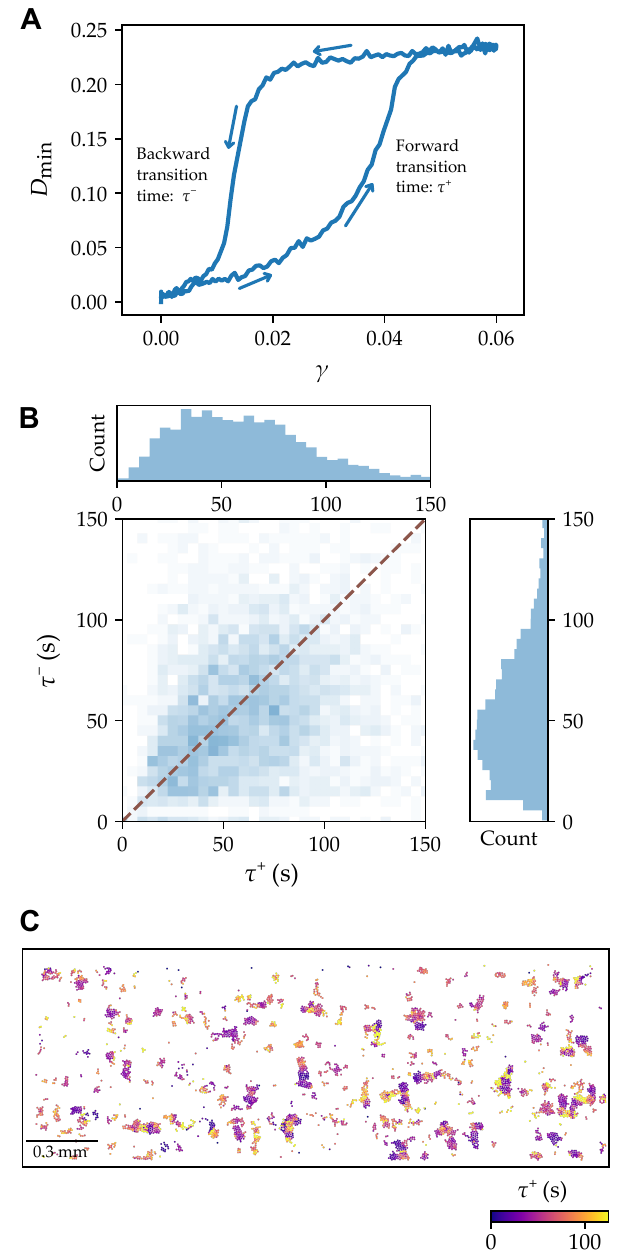}
    \caption{
    \textbf{({A})} Plot of $D_{\mathrm{min}}$ of one particle against shear strain $\gamma$ during one cycle. $D_{\mathrm{min}}$ measures the non-affine displacement of a particle compared to its initial placement among its nearest neighbors. Arrows mark the direction of travel, starting at lower left. In this example, the local structure is bistable in the range of shear strain $0.014<\gamma<0.038$, which means its state in this range depends on its history. 
    \textbf{({B})} Distribution of rearrangement timescales $(\tau^+$, $\tau^-)$ (measured in seconds), for {4161} rearranging particles at $f=0.004$~Hz. Central plot: Darker color indicates greater frequency. Dashed diagonal line indicates $\tau^+=\tau^-$. Histograms on axes: sums along each row/column. 
    \textbf{({C})} Spatial map of the rearranging particles in (B), colored by rearrangement timescale $\tau^+$. Each dot marks the location of one particle (not indicative of size). Particles in the same rearranging area tend to have similar relaxation times. %
    }
    \label{fig:dmin}
\end{figure}

Figure~\ref{fig:dmin}(A) shows $D_{\min}$ of a particle at the center of one rearranging region, as a function of the global shear strain $\gamma$. The curve is single-valued near the turning points of the motion, but at intermediate strains it shows bistability and hysteresis. Although this loop resembles the magnetization of a ferromagnet as a function of the external field, in that system the response comprises many magnetic domains with a wide distribution of properties, whereas here the curve arises from the coordinated particle motions within a single rearrangement. Conceptually, the area within the hysteresis loop is associated with the elastic potential energy that is lost to viscous drag when particles rearrange. The symmetric shape of a typical $D_{\min}$ curve is noteworthy, given that unlike magnetization, $D_{\min} \ge 0$ and it is measured relative to an extremum of the motion. We obtain nearly the same curve (up to a reflection in $D_\text{min}$) when choosing the reference configuration at $\gamma = \gamma_\text{max}$ or any other strain where the curve is nearly flat. 

For particles in the same rearrangement, the shapes of the particles' $D_{\min}$ curves are usually similar, owing to their coordinated motion. The $D_{\min}$ of a single particle is therefore a one-dimensional proxy for the kinematics of its rearranging region. In the rest of our analysis, we use the $D_{\min}$ of single particles to study the dynamics of rearrangements.

In Fig.~\ref{fig:dmin}(A), the forward ($\dot \gamma > 0$) and reverse ($\dot \gamma < 0$) transitions between high and low $D_\text{min}$ have characteristic rates in time. 
The corresponding timescales $\tau^+$ and $\tau^-$ are estimated using the elapsed time between 40\% and 60\% of the maximum $D_{\mathrm{min}}$, which we then extrapolate to the whole transition between 0\% and 100\% by dividing over $(60\%-40\%)$. This avoids the noise around 0\% and 100\% maximum $D_{\mathrm{min}}$. The histogram for all particles is shown in Fig.~\ref{fig:dmin}. 
The timescales span more than a decade, with a significant population near the half period of the cycle ($T/2 = 129$~s)---effectively the maximum time available for a transition in this experiment. The spread away from $\tau^+=\tau^-$ indicates that the timescales are only weakly correlated, with some overall asymmetry that is possibly due to a global directional memory~\cite{Galloway2021,Keim2022,Adhikari2022} or the 5~s pause after each cycle.

While Fig.~\ref{fig:traj}(B) shows a broad distribution of timescales, we would expect particles participating in the same local rearrangement event to have similar timescales. In Fig.~\ref{fig:dmin}(C) we plot the position of all 4161 rearranging particles, and color them with their respective $\tau^+$. Indeed neighboring particles tend to have similar rearranging timescales, indicating a collective motion with a characteristic length scale of about 4 or 5 interparticle distances. In larger regions several timescales may be present, sometimes with clear boundaries between different subregions; in other cases there is less distinction. Timescales are not strongly correlated with cluster sizes or the distance to the boundaries. Consistent with the histogram, timescales longer than 100~s are less common. They tend to appear in smaller clusters, and are often adjacent to other clusters. The very smallest clusters, containing 1 or 2 particles, are due to tracking errors, or represent marginal cases where the amplitude of the displacement is close to the detection threshold. %

\begin{figure}
    \centering
    \includegraphics[width=0.48\textwidth]{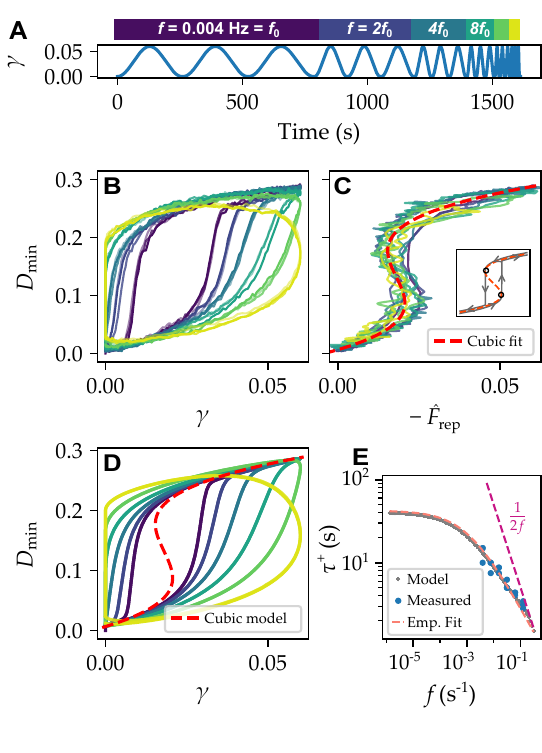}
    \caption{Observed rearrangement dynamics depend on frequency.
    \textbf{(A)} Frequency sweep protocol, increasing by factors of 2 from $f=f_0 = 0.004$~Hz to $f=32 f_0$.
    \textbf{(B)} $D_{\mathrm{min}}$ trajectory for a single particle during the frequency sweep. The three cycles at each frequency use the color bar above (A). %
    Repetitions have the same color but diminished intensity.
    \textbf{(C)} Using Eq.~\eqref{eqn:force2}, we obtained the $D_{\min}$-dependent $\hfrep$ curve, and show that it is independent of driving frequency and direction. %
    Loops at the same frequencies are averaged over.
    Dashed line is the best fit cubic force model. Inset: interpreting the $\hfrep$ curve as the quasi-static limit of the rearrangement; stable portions of the curve are followed up to the limits of stability. %
    \textbf{(D)} Model output using identical driving waveform and force model. 
    \textbf{(E)} Extracted timescales from (B), compared with model output with the same force model and variable frequency. The model is extended by over 3 decades in the low frequency range, showing the transition from power-law to quasi-static regime. Orange dashed line shows an empirical fit (see text). %
    $1/(2f)$ (purple dashed line) bounds $\tau^+$ from above. %
    }
    \label{fig:theory}
\end{figure}

We also observe dynamics as we vary the timescale of driving. In previous work on structure~\cite{Keim2013,Keim2014,Keim2015,Galloway2021} and memory~\cite{Keim2020,Keim2022} in this system, the lowest frequency was 0.05~Hz---more than an order of magnitude faster than the base frequency $f_0 = 0.004$~Hz we have used so far. We double the frequency 5 times, up to 0.124~Hz, repeating each frequency for 3 cycles. The strain protocol is shown in Fig.~\ref{fig:theory}(A). 
In Fig.~\ref{fig:theory}(B) we plot $D_\text{min}$ vs.\ $\gamma$ for a typical rearrangement during the entire protocol, using the $D_{\min}$ curve from one of the particles near the center of the rearrangement where the non-affine displacement is maximal. We find that as frequency increases, each forward rearrangement appears to occur at a larger strain (measured at 50\% of maximum $D_\textrm{min}$), and each reverse rearrangement appears to occur at a smaller strain, widening the observed hysteresis loop. In higher-frequency cycles, part of the backwards transition overlaps with the 5~s waiting period at $\gamma=0$. This appears in Fig.~\ref{fig:theory}(B) as a vertical line going downwards at $\gamma=0$.

\subsection{One-dimensional model}

Treating the particle's non-affine displacement $D_\text{min}$ as a coordinate variable lets us seek a one-dimensional model for the observed behaviors. Because the inertia of the particle is negligible, the equation of motion for $D_\text{min}$ is the force balance
\begin{equation}\label{eqn:force}
\eta\frac{dD_{\min}}{dt}=\frep(D_{\min})+F_D\gamma(t)
\end{equation}
where $\eta$ acts as a linear drag coefficient, $\frep$ is the position-dependent effective force due to repulsion by a particle's neighbors, and $F_D\gamma(t)$ is a driving force that is proportional to the imposed strain.
We non-dimensionalize the equation of motion as
\begin{equation}\label{eqn:force2}
\gamma(t)-\hat\eta\frac{dD_{\min}}{dt}=-\hfrep(D_{\min})
\end{equation}
where $\hfrep=\frep/F_D$, $\hat \eta=\eta/F_D$. %
Note that Eq.~\eqref{eqn:force2} does not explicitly depend on time; instead, it depends on $D_{\min}(t)$ as $\hfrep$ does. 
On the left-hand side, $\gamma(t)$ and $dD_{\min}/dt$ are experimentally accessible quantities, and $\hat\eta$ is a free fitting parameter. Finding the appropriate value of $\hat\eta$ should let us subtract the drag term, removing the dependence on $dD_{\min}/dt$ and revealing $\hfrep(D_{\min})$. Because $\hfrep(D_{\min})$ should be single-valued, finding $\hat\eta$ is equivalent to minimizing the loop area in Fig.~\ref{fig:theory}(C). We obtain $\hat\eta\approx 0.7$~s (see SI~\cite{supp} for details). This best-fit value is consistent with the $\hat \eta \sim 1$~s expected for viscous drag on a particle with typical interparticle forces (see SI~\cite{supp}).
The resulting collapsed curve is largely independent of frequency and driving direction, supporting our model assumptions. 

The $-\hfrep$ values in Fig.~\ref{fig:theory}(C) represent all static equilibrium states (i.e.\ where $\gamma = -\hfrep$ in Eq.~\ref{eqn:force2}). The portions with positive slope represent static equilibria that are stable when $\gamma$ is held fixed. The most important feature is the folded-over portion, where one $\hfrep$ value corresponds to two stable $D_{\min}$ values. At each boundary of this portion, one branch of stable equilibria ends and $D_{\min}$ switches to the other stable branch; returning to the first branch involves hysteresis (Fig.~\ref{fig:theory}(C) inset). 

To capture this behavior, we approximate $\hfrep(D_{\min})$ with a cubic curve
\begin{equation}\label{eqn:force3}
\hfrep = -\hat A(D_{\min}-\hat x_b)^3+\hat B(D_{\min}-\hat x_b)+\hat F_P
\end{equation}
Fitting the curve to the data gives the nondimensional model parameters $\hat A$, $\hat B$, $\hat x_b$ and $\hat F_P$. 
When $\hat A, \hat B>0$, the cubic force corresponds to a potential energy that is a quartic function of $D_\text{min}$, with local minima at 
$\hat x_b\pm\sqrt{\hat B/\hat A}\equiv \hat x_b\pm \hat x_0$, separated by a barrier; this is given by
\begin{multline}\label{eqn:pot}
\hat U(D_{\min})=\frac14\hat A(D_{\min}-\hat x_b)^4-\frac12\hat B(D_{\min}-\hat x_b)^2\\-\hat F_P(D_{\min}-\hat x_b)
\end{multline}
up to a constant offset. Crossing the barrier involves a characteristic force $\hat F_0=(2/3)\sqrt{\hat B^3/(3\hat A)}$, while 
$\hat F_P$ acts as a bias that gives these potential minima different depths. The resulting limits of stability are at $(\hat F_P \pm \hat F_0, \hat x_b\mp\hat x_0/\sqrt{3})$ in the coordinates of Fig.~\ref{fig:theory}(C), marked by circles in the inset figure. 
Our model is closely related to the dynamical hysteresis concept that has been applied to ferromagnetic systems and bistable lasers~\cite{Rao1990,Jung1990}. 
In the zero-frequency limit, $\hat F_P\pm\hat F_0$ become the strain thresholds $\gamma^\pm$ at which the system leaves one stable branch and switches to the other (Fig.~\ref{fig:theory}(C) inset), approximating the ``hysteron'' model of a bistable unit that switches instantaneously \cite{Preisach1935,Keim2019,Mungan2019}. 
If driving is held constant (i.e.\ zero strain rate), the parameters give a relaxation timescale 
$\tau_0=\hat \eta / \hat B$ for reaching a stable minimum. %

Using the fit parameters obtained above, we integrate Eq.~\eqref{eqn:force2} using the experimental $\gamma(t)$ waveform. As shown in Fig.~\ref{fig:theory}(D), the model captures the frequency dynamics of the rearrangement very well, with the loops widening as frequency increases, corresponding to greater dissipation per cycle. 
Remarkably, at the highest frequency the peak $D_\text{min}$ is slightly smaller and occurs near $\gamma=0$, well after the global deformation has reversed.

\subsection{Apparent timescales depend on frequency}

The frequency sweep experiment reveals that the apparent timescales $\tau^\pm$ in Fig.~\ref{fig:dmin} are not fixed properties, but increase as frequency increases. 
In Fig.~\ref{fig:theory}(E), we plot the $\tau^+$ values measured during the experimental frequency sweep, which agree well with a smooth curve of measurements from the model. Both show an approximate $f^{-\beta}$ scaling in the range of frequencies probed, with $\beta\approx 0.6$ from numerical solutions (see SI~\cite{supp}). An empirical form of the simulation curve can be written as 
\begin{equation}
{\tau^+}\approx\frac{\tau_{QS}}{1+(f/f_C)^{0.6}}
\label{eqn:scaling}
\end{equation}
where $\tau_{QS}$ and $f_C$ are fitting parameters whose significance is discussed below. A rigorous derivation of Eq.~\eqref{eqn:scaling} is beyond the scope of this work, although it may follow a scaling argument similar to those in Refs.~\citenum{Jung1990, Broner1997}.

In the limit of low frequency and strain rate, $\gamma$ stays nearly fixed while particles rearrange, and we expect the observed $\tau^\pm$ to approach constant values $\sim \tau_0$. Extrapolating our model in Fig.~\ref{fig:theory}(E), we observe that this quasi-static limit with constant timescale is not reached for $\sim$2 decades below the lowest frequency in our experiment, corresponding to a period of several hours. In this limit, numerical integration gives $\tau^+ \to \tau_{QS}$. 

To estimate $\tau_{QS}$ analytically, we take the limit $f \to 0$, and use the approximation $\tau^+\approx\max(D_{\min})/(dD_{\min}/dt)=\max(D_{\min})\hat\eta/(\hfrep+\gamma)$. In the $f \to 0$ limit, the transition takes place at $\gamma=\gamma^+$, and it follows that $\hfrep+\gamma^+=\hat F_0$, which is the characteristic force. Therefore, we estimate $\tau_{QS}\sim\max(D_{\min})\hat\eta/\hat F_0= (3/2)\sqrt{3} \tau_0\max(D_{\min})/\hat x_0$. For the parameters corresponding to Fig.~\ref{fig:theory} this gives $\tau_{QS}\approx50$~s, consistent with the numerical result in Fig.~\ref{fig:theory}(E). 

At frequencies above this quasi-static regime, $\gamma$ is no longer constant during the transition; to the next order, the timescale depends on the driving strain rate. The crossover frequency is where the strain rate and the rate of $\hfrep$ relaxation are comparable, $d\gamma/d t \approx \hat F_0/\tau_0$. Taking $d\gamma/d t \approx 2\pi f(\gamma_{\max}/2)$, 
this gives the crossover frequency  $f_C = \hat F_0 / (\pi \gamma_\mathrm{max} \tau_0 ) \approx 2 \times 10^{-3}$~Hz between the quasi-static and scaling regimes. The quasi-static regime requires that the strain rate is negligible, implying $f \ll f_C$, consistent with numerical results (Fig.~\ref{fig:theory}(E)). 

At high frequencies, $\tau^\pm$ must be bounded from above by the half-period of driving, and we find that the scaling regime (Eq.~\ref{eqn:scaling}) breaks down near the intersection with $1/(2f)$. The subsequent regime is not evident in Fig.~\ref{fig:theory}, but we will examine it in Sec.~\ref{sec:skipping}.

Our results mean that over a wide range of frequencies, the observed transition timescales $\tau^\pm$ are not intrinsic properties of a rearrangement. 
Instead, as the period of driving grows longer, the timescales increase sub-linearly with inverse frequency, and saturate at a quasi-static limit. 
Intuitively, one might suppose that the quasi-static limit is reached when the apparent timescale is significantly shorter than the period of driving, e.g.\ $\tau^+ \approx 10^{-1} f^{-1}$. In practice, Fig.~\ref{fig:theory}(E) shows that in this condition, the observed $\tau^+$ will be only $\sim$$1/4$ of its quasi-static value. To actually observe $\tau^\pm \approx \tau_{QS}$, one must continue lowering the frequency until $\tau^+ \lesssim 10^{-2} f^{-1}$, i.e.\ $f \lesssim 10^{-4}$~Hz. This counter-intuitive slowness comes from the nature of instability in an overdamped system: just when the system loses stability, the forces of driving and repulsion on the particle are nearly balanced. In dynamical-systems terms, the $D_{\min}$ values where a rearrangement loses stability are saddle-node remnants, near which the dynamics becomes stagnant~\cite{Strogatz2018}. The system will stay in this region for a long time ($\sim (\gamma-\gamma^+)^{-1/2}$ in the static case) unless the driving force $\gamma$ changes significantly. %
Therefore, despite the small size of these saddle-node remnants in $D_{\min}$, they play a disproportionate role in the extraordinary slowness of $\tau^\pm$ at low frequencies. 

Given that the particle in Fig.~\ref{fig:theory}(C) is far from quasi-static at experimental frequencies, we may ask whether any particles in our experiment do show dynamics in the low-frequency limit. Noting the low-frequency plateau in Fig.~\ref{fig:theory}(E), we sought particles for which $\tau^+$ measured at the two lowest frequencies is near the predicted $\tau_{QS}$ from $\tau_0$ and $\hat x_0$. %
No particle's $\tau^+$ closely matches the predicted $\tau_{QS}$ within 30\%, suggesting that the quasi-static regime lies well below 0.004~Hz. 
The transitions that come closest to $\tau_{QS}$ are ones that happen within 10\% of $\gamma_{\max}$, when the strain rate is smallest, emphasizing that the amplitude of driving can also affect dynamics---a point we revisit in Sec.~\ref{sec:skipmodel}. %

\begin{figure}
    \centering
    \includegraphics[width=\linewidth]{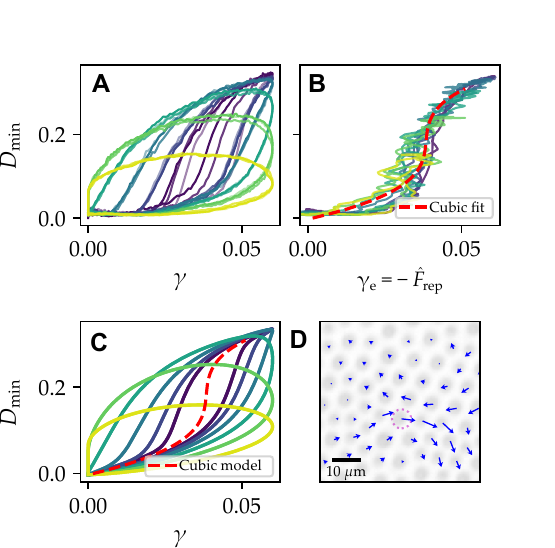}
    \caption{Example of rearrangement-like motion without static hysteresis. 
    \textbf{(A)} Frequency-dependent $D_{\min}$ loops. 
    \textbf{(B)} Inferred $\hfrep$ curve; showing positive $\hat B$ term. As such, no hysteresis exists in the low frequency limit. 
    \textbf{(C)} Model output from integrating Eq.~\eqref{eqn:force2}. 
    \textbf{(D)} Particle non-affine displacements in half a cycle, similar to those of Fig.~\ref{fig:traj}(B), for the rearrangement shown in (A--C) corresponding to the lowest frequency; circle marks the particle whose $D_{\min}$ is measured. Arrows are twice the actual displacement for clarity.}
    \label{fig:nonhysteretic}
\end{figure}

\subsection{Monostable rearranging motion}

By extracting the effective one-dimensional force, we find that a small number of rearranging regions look superficially like the rest, but have underlying physics that are distinct. We noted above that at low frequency, a rearrangement becomes more like the hysteron model, with threshold strains $\gamma^\pm$ given by the limits of stability in the $\hfrep(D_{\min})$ curve. However, this is only true when the parameters $\hat A,\,\hat B>0$~\cite{Broner1997}. In Fig.~\ref{fig:nonhysteretic} we show an example where $\hat A>0$ and $\hat B<0$. The magnitudes of the displacements and their extended quadrupolar pattern are similar to those of Fig.~\ref{fig:traj}, and the motion is hysteretic with respect to strain. However, the hysteresis is due to dynamics alone: the loop shrinks with decreasing $f$, eventually approaching the static $-\hfrep$ curve (given in Fig.~\ref{fig:nonhysteretic}(B)) which is strongly anharmonic without also being bistable. We found just two such regions in our sample, suggesting that the behavior is relatively rare.

The meaning and mechanism of the monostable behavior are unclear. The central portion of the force curve in Fig.~\ref{fig:nonhysteretic}(B) is nearly vertical, suggesting an exceptionally weak restoring force. In principle these particle motions could contribute to low-frequency vibrational properties without contributing to plasticity. We note that Szulc et al.~\cite{Szulc2022} observed in molecular dynamics simulations that in rare cases, a bistable rearrangement seemed to become monostable due to the influence of another, nearby rearrangement. %

\begin{figure}
    \centering
    \includegraphics[width = 0.48\textwidth]{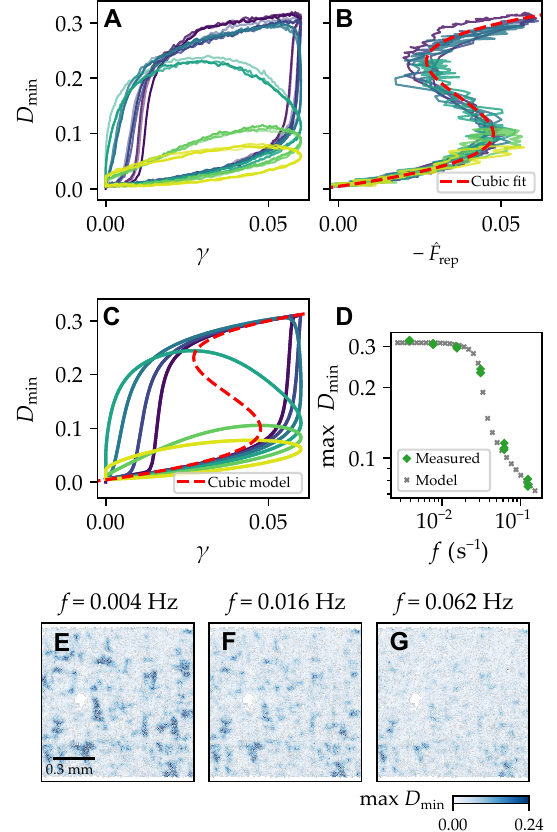}
    \caption{Because of their dynamics, some rearrangements can be skipped at high frequencies.
    \textbf{(A)} $D_{\mathrm{min}}$ loops for a central particle in the rearrangement of Fig.~\ref{fig:traj}(A--C), showing a sharp decrease in the non-affine motion as driving frequency increases.
    \textbf{(B)} The same curves as in (A) but with frequency dependent parts removed, as before. Notice that the rearrangement is stalled at the turning point for the highest two frequencies.
    \textbf{(C)} Model output with identical driving waveform and force model. 
    \textbf{(D)} The peak $D_\mathrm{min}$ drops quickly at a threshold frequency $\sim 0.03$~s$^{-1}$, which is verified with model output. 
    \textbf{(E--G)} Plot of $D_{\min}$ in a representative region, for different driving frequencies.
    }
    \label{fig:skipping}
\end{figure}

\subsection{Skipping at high frequency}
\label{sec:skipping}

In Fig.~\ref{fig:traj}, we showed that a rearrangement can be skipped at higher frequencies, so that the material's local response is elastic. 
To investigate, in Fig.~\ref{fig:skipping} we analyze a central particle from the rearranging area in Fig.~\ref{fig:traj}(A). The low frequency cycles for $f\le 0.016$ Hz consistently reach the peak $D_{\min}$ value near the turning point of the deformation, $\gamma = \gamma_\text{max}$, which we interpret as the rearrangement being completed. As in Fig.~\ref{fig:theory}(B), the transitions are progressively more delayed as frequency increases. At $f=0.031$ Hz, the forward rearrangement seems to begin too late to reach completion before the driving strain is reversed, and the peak $D_{\min}$ is delayed so much that it occurs closer to $\gamma=0$ than to $\gamma = \gamma_\text{max}$. At still higher frequencies, the rearrangement is skipped, with highest $D_{\min}$ much less than 50\% of the previous peak values.
We performed the same analysis as before, shown in Fig.~\ref{fig:skipping}(B, C). Most notably, the inferred force curve shows an unstable equilibrium at the bend near $(\gamma,D_{\min})=(0.05,0.1)$, which is close to $\gamma_{\max}$. Thus $\gamma$ cannot grow large enough to quickly drive the system past the unstable equilibrium, creating the possibility that the particles stay in the configuration with low $D_{\min}$. 

We quantify skipping with the maximum $D_{\min}$ attained during a cycle, which we plot as a function of $f$ in Fig.~\ref{fig:skipping}(D). A sharp drop occurs at a threshold frequency $\sim 0.04$ Hz. At the highest frequency, the maximum $D_{\min}$ value drops below 0.1, which is the static limit of stability. As such, our model lets us conclude that the rearrangement was suppressed completely, and the trajectories at the highest frequency represent driven oscillations around a stable equilibrium. The model output is consistent with the skipping picture. The driving frequency alters not only the dynamics of how particles rearrange, but whether they rearrange at all.

How common are skipped rearrangements? Although rearrangements have different timescales and limits of stability, we expect that when strain amplitude is held constant, each rearrangement will skip if driven above some frequency. In Fig.~\ref{fig:skipping}(E--G) we show a section from the region in Fig.~\ref{fig:dmin}(C), with particles colored by the maximum $D_{\min}$ values attained at each frequency. 
Between 0.004~Hz and 0.016~Hz, most rearrangements remain in place though the magnitudes are reduced. At 0.062~Hz most rearrangements vanish. Overall, nonaffine displacements are weaker and more localized than at low frequency. This progression is reminiscent of rate-dependent experiments in concentrated emulsions, where the characteristic length of nonaffine regions was found to decrease with frequency~\cite{Vasisht2018}, as well as computational studies of brittle yielding in amorphous solids with damping~\cite{Singh2020}, which may indicate a similar viscous origin of these effects.

\subsection{Skipping criterion in parameter space}
\label{sec:skipmodel}

Within the model's wide range of parameters and behaviors, we now focus on the essential question posed by Fig.~\ref{fig:traj}: whether driving will cause a rearrangement to complete, or to be skipped. For simplicity, we show results from integrating the non-dimensional equation 
\begin{equation}\label{eqn:1}
    dX/dT=X-X^3-A\cos(\Omega T)-C
\end{equation}
which is obtained from rescaling Eqs.~\ref{eqn:force}--\ref{eqn:force3}, under the transforms $X=(D_\textrm{min}-\hat x_b)/\hat x_0$, $T=t/\tau_0$, $A=\gamma_{\max}/\hat F_0$, $C=\hat F_P/\hat F_0+\gamma_{\max}/2$. This restricts the independent parameters to driving amplitude $A$, driving frequency $\Omega$, and the asymmetry or bias of the force, $C$, from the original $\{\hat\eta, \hat A, \hat x_b, \hat B, \hat F_P, \gamma_{\max}, f\}$. 
Any solution to Eq.~\ref{eqn:1} can be brought into the form of Eqs.~\ref{eqn:force}--\ref{eqn:force3} through simple translation and rescaling operations. We further restrict our analysis to the simplest case, $C = 0$. In the potential-energy representation, this scenario corresponds to a quartic potential with two identical minima at $X = \pm 1$.

While transient solutions to Eq.~\eqref{eqn:1} are interesting and can arise in experiments, here we focus on limit cycles. For the limit cycle $X(T)$, we define the order parameter $\bar{X}$ to be the average of $X$ over one cycle, i.e.\ $\Omega/(2\pi)\int_0^{2\pi/\Omega}X(T')dT'$. When $A > 1$, the low-frequency ($\Omega \ll 1$) limit cycle has $\bar X=0$, since $X$ can switch between the two stable branches. In the limit $\Omega\to\infty$ this switching is suppressed and $\bar X = \pm 1$; the limit cycle is an oscillation around a stable equilibrium. Without loss of generality, we chose the initial condition that corresponds to the upper branch, giving $\bar X=1$. 

\begin{figure}
    \centering
    \includegraphics[width=\linewidth]{phasediagram.pdf}
    \caption{Characterizing the steady state solutions of the dynamical equation. $\bar X$ is the cycle-average of $X$. The parameter space can be divided into two regions: one corresponds to periodically switching solutions where $X$ moves between two stable branches, with $\bar X=0$ since the model is symmetric; the other is the trapped solution where $\bar X$ is attracted to one of the wells at $X=\pm1$ (for simplicity, only the $X=+1$ branch is shown); which one it is attracted to depends on history. The two phases are separated by a transitional regime near $\Omega\sim A-1$.}
    \label{fig:phase_diagram}
\end{figure}

The numerical results in Fig.~\ref{fig:phase_diagram} show that between these limiting cases, the limit-cycle behavior remains surprisingly simple. 
The line of separation is somewhere between $\Omega = A-1$ and $\Omega = (A-1)^{0.8}$; this transition is sharp for small values of $A-1$ and $\Omega$, and becomes smoother for large values. Therefore, by either increasing frequency or decreasing driving amplitude (or a combination of both), one can bring the system out of the periodic switching regime, causing it to remain near one of the stable energy minima as in Fig.~\ref{fig:skipping}. Because the system is bistable, which potential well it dwells in depends on the initial condition and the driving history.

\section{Discussion and Conclusion}

Our experiments studied the dynamics of plastic rearrangements that are caused by repeatedly shearing a soft, athermal jammed solid. Viscous drag makes these dynamics observable in our colloidal solid, and oscillatory shear makes them prominent. We find that the apparent timescale of a rearrangement, and the strain values at which it appears to switch configurations, depend on the frequency of driving. Furthermore, while we find that at a given frequency most rearrangements proceed with a similar timescale, there is a significant minority with rates up to 10 times slower, near the slowest we can observe. Because raising the driving frequency can cause some rearranging regions to respond elastically, it is likely that further lowering the frequency could reveal latent rearrangements. Thus even with a driving period of 258~s, it is unlikely that our experiments have observed the slowest timescales in our sample, nor the full set of plastic rearrangements that could be activated within our limits of strain. 
Our results may be connected to other, known forms of surprising slowness in glass and glassy systems, including the numerous quasi-localized low-frequency vibrational or relaxational modes~\cite{Lerner2021} that seem to be associated with plastic rearrangements~\cite{Manning2011}, and the phenomenon of logarithmic aging in which rearrangements become increasingly rare but never seem to stop~\cite{Amir2012,Shohat2025}. 

While the motions of rearranging particles are two-dimensional and complex (Fig.~\ref{fig:traj}(A, B)), they are coordinated. We found that the motion can be well approximated with one degree of freedom (Fig.~\ref{fig:theory}(C, D)). In this simplified representation, we can obtain a frequency-independent effective potential energy (Eq.~\ref{eqn:pot}) or force (Eq.~\ref{eqn:force3}) that depends only on one coordinate (here, $D_{\min}$). This potential typically has a double-well (quartic) form, and with it we can reproduce the behavior of a rearrangement over the entire range of driving frequencies. 
However, the same results show that non-affine particle motions do \emph{not} inherently indicate plastic rearrangements: for each particle we analyze (Figs.~\ref{fig:theory}(D), \ref{fig:nonhysteretic}(C),  and \ref{fig:skipping}(D)), $D_\text{min}$ can change significantly while the particle is still stable. In rare cases, a rearrangement remains stable even as particles move by distances that are typical for bistable rearrangements (Fig.~\ref{fig:nonhysteretic}). Future work may clarify the origin of these ``monostable'' motions and assess their prevalence across a wider range of systems. There is a particular need for molecular dynamics simulations that detect rearrangements via particle motions, rather than sharp energy drops. %

Since viscous damping is a common feature of soft solids made of colloids, hydrogel beads, bubbles, or droplets, our results suggest that comparisons with quasistatic models and simulations (which approximate zero frequency and strain rate) must be made with caution. Of particular concern is the assumption that the strain thresholds at which a rearranging region becomes unstable (often denoted $\gamma^\pm$) are independent of driving, and that a rearrangement will be triggered if one of its thresholds falls between the turning points of driving. 
Even an apparently low-frequency experiment can deceive: when one compares the period of driving to the observed timescales of rearrangements, a decade of separation (i.e.\ Deborah number $\sim$$10^{-1}$) would customarily suggest that the frequency is sufficiently low. However, results like Fig.~\ref{fig:theory}(E) mean that the actual low-frequency regime may still be far away.
While we cannot compare our results directly to prior experiments with this system~\cite{Keim2013,Keim2014,Keim2020, Keim2022} because the area fraction is different, our results clearly show that apparent timescales can be misleading when assessing dynamics.
The gap between experiments and the low-frequency limit may be especially acute in studies of mechanical annealing, yielding, and irreversibility:
the number density of rearrangements may always be lower at finite frequency, reducing the possibility that nearby rearrangements will overlap and become irreversible~\cite{Szulc2024} or form a nascent shear band~\cite{Jagla2007,Martens2012}.

Our results raise practical challenges and exciting opportunities for future experiments. We have shown that in a system with linear damping, particle trajectories can be used to reconstruct the dynamics of plastic rearrangements, and extract frequency-independent properties---but only for rearrangements that are fast enough to be observed. This prompts us to ask how the dynamical parameters we have uncovered are evident in static structure, especially given recent successes in predicting the local propensity to rearrange~\cite{Cubuk2017,Richard2020}. The finding that strain thresholds shift and rearrangements skip as a function of frequency also opens the possibility of dynamical memory behaviors that count cycles after the startup of driving~\cite{Lindeman2023} or reach states that would be inaccessible at low frequencies~\cite{Jules2023,Gutierrezprieto2025}---effectively, memories of full dynamical trajectories rather than the turning points of driving. %

\begin{acknowledgments}
    We thank Doug Durian, Chloe Lindeman, Craig Maloney, Sidney Nagel, and Arjun Yodh for helpful conversations. This work was supported in part by grant DMR-2346012 from the National Science Foundation.
\end{acknowledgments}

\nocite{Danov2010,Masschaele2010,Chen2005,Tajuelo2016,Reynaert2008} %

\bibliographystyle{apsrev4-1}
\bibliography{export1, export2}

\begin{thebibliography}{44}%
\makeatletter
\providecommand \@ifxundefined [1]{%
 \@ifx{#1\undefined}
}%
\providecommand \@ifnum [1]{%
 \ifnum #1\expandafter \@firstoftwo
 \else \expandafter \@secondoftwo
 \fi
}%
\providecommand \@ifx [1]{%
 \ifx #1\expandafter \@firstoftwo
 \else \expandafter \@secondoftwo
 \fi
}%
\providecommand \natexlab [1]{#1}%
\providecommand \enquote  [1]{``#1''}%
\providecommand \bibnamefont  [1]{#1}%
\providecommand \bibfnamefont [1]{#1}%
\providecommand \citenamefont [1]{#1}%
\providecommand \href@noop [0]{\@secondoftwo}%
\providecommand \href [0]{\begingroup \@sanitize@url \@href}%
\providecommand \@href[1]{\@@startlink{#1}\@@href}%
\providecommand \@@href[1]{\endgroup#1\@@endlink}%
\providecommand \@sanitize@url [0]{\catcode `\\12\catcode `\$12\catcode `\&12\catcode `\#12\catcode `\^12\catcode `\_12\catcode `\%12\relax}%
\providecommand \@@startlink[1]{}%
\providecommand \@@endlink[0]{}%
\providecommand \url  [0]{\begingroup\@sanitize@url \@url }%
\providecommand \@url [1]{\endgroup\@href {#1}{\urlprefix }}%
\providecommand \urlprefix  [0]{URL }%
\providecommand \Eprint [0]{\href }%
\providecommand \doibase [0]{http://dx.doi.org/}%
\providecommand \selectlanguage [0]{\@gobble}%
\providecommand \bibinfo  [0]{\@secondoftwo}%
\providecommand \bibfield  [0]{\@secondoftwo}%
\providecommand \translation [1]{[#1]}%
\providecommand \BibitemOpen [0]{}%
\providecommand \bibitemStop [0]{}%
\providecommand \bibitemNoStop [0]{.\EOS\space}%
\providecommand \EOS [0]{\spacefactor3000\relax}%
\providecommand \BibitemShut  [1]{\csname bibitem#1\endcsname}%
\let\auto@bib@innerbib\@empty
\bibitem [{\citenamefont {Rubinstein}\ and\ \citenamefont {Colby}(2003)}]{rubinstein}%
  \BibitemOpen
  \bibfield  {author} {\bibinfo {author} {\bibfnamefont {M.}~\bibnamefont {Rubinstein}}\ and\ \bibinfo {author} {\bibfnamefont {R.~H.}\ \bibnamefont {Colby}},\ }\href {\doibase 10.1093/oso/9780198520597.001.0001} {\emph {\bibinfo {title} {Polymer Physics}}}\ (\bibinfo  {publisher} {Oxford University Press},\ \bibinfo {year} {2003})\BibitemShut {NoStop}%
\bibitem [{\citenamefont {Bandi}\ \emph {et~al.}(2018)\citenamefont {Bandi}, \citenamefont {E.~Hentschel}, \citenamefont {Procaccia}, \citenamefont {Roy},\ and\ \citenamefont {Zylberg}}]{Bandi2018}%
  \BibitemOpen
  \bibfield  {author} {\bibinfo {author} {\bibfnamefont {M.~M.}\ \bibnamefont {Bandi}}, \bibinfo {author} {\bibfnamefont {H.~G.}\ \bibnamefont {E.~Hentschel}}, \bibinfo {author} {\bibfnamefont {I.}~\bibnamefont {Procaccia}}, \bibinfo {author} {\bibfnamefont {S.}~\bibnamefont {Roy}}, \ and\ \bibinfo {author} {\bibfnamefont {J.}~\bibnamefont {Zylberg}},\ }\href@noop {} {\bibfield  {journal} {\bibinfo  {journal} {EPL (Europhysics Letters)}\ }\textbf {\bibinfo {volume} {122}},\ \bibinfo {pages} {38003} (\bibinfo {year} {2018})}\BibitemShut {NoStop}%
\bibitem [{\citenamefont {Shohat}\ \emph {et~al.}(2025)\citenamefont {Shohat}, \citenamefont {Baconnier}, \citenamefont {Procaccia}, \citenamefont {Hecke},\ and\ \citenamefont {Lahini}}]{Shohat2025}%
  \BibitemOpen
  \bibfield  {author} {\bibinfo {author} {\bibfnamefont {D.}~\bibnamefont {Shohat}}, \bibinfo {author} {\bibfnamefont {P.}~\bibnamefont {Baconnier}}, \bibinfo {author} {\bibfnamefont {I.}~\bibnamefont {Procaccia}}, \bibinfo {author} {\bibfnamefont {M.~v.}\ \bibnamefont {Hecke}}, \ and\ \bibinfo {author} {\bibfnamefont {Y.}~\bibnamefont {Lahini}},\ }\href@noop {} {\bibfield  {journal} {\bibinfo  {journal} {arXiv}\ ,\ \bibinfo {pages} {2506.08779v1}} (\bibinfo {year} {2025})}\BibitemShut {NoStop}%
\bibitem [{\citenamefont {Falk}\ and\ \citenamefont {Langer}(1998)}]{Falk1998}%
  \BibitemOpen
  \bibfield  {author} {\bibinfo {author} {\bibfnamefont {M.~L.}\ \bibnamefont {Falk}}\ and\ \bibinfo {author} {\bibfnamefont {J.~S.}\ \bibnamefont {Langer}},\ }\href {\doibase 10.1103/PhysRevE.57.7192} {\bibfield  {journal} {\bibinfo  {journal} {Physical Review E}\ }\textbf {\bibinfo {volume} {57}},\ \bibinfo {pages} {7192} (\bibinfo {year} {1998})}\BibitemShut {NoStop}%
\bibitem [{\citenamefont {Falk}\ and\ \citenamefont {Langer}(2011)}]{Falk2011}%
  \BibitemOpen
  \bibfield  {author} {\bibinfo {author} {\bibfnamefont {M.~L.}\ \bibnamefont {Falk}}\ and\ \bibinfo {author} {\bibfnamefont {J.}~\bibnamefont {Langer}},\ }\href {\doibase 10.1146/annurev-conmatphys-062910-140452} {\bibfield  {journal} {\bibinfo  {journal} {Annual Review of Condensed Matter Physics}\ }\textbf {\bibinfo {volume} {2}},\ \bibinfo {pages} {353} (\bibinfo {year} {2011})}\BibitemShut {NoStop}%
\bibitem [{\citenamefont {Lindeman}\ \emph {et~al.}(2023)\citenamefont {Lindeman}, \citenamefont {Hagh}, \citenamefont {Ip},\ and\ \citenamefont {Nagel}}]{Lindeman2023}%
  \BibitemOpen
  \bibfield  {author} {\bibinfo {author} {\bibfnamefont {C.~W.}\ \bibnamefont {Lindeman}}, \bibinfo {author} {\bibfnamefont {V.~F.}\ \bibnamefont {Hagh}}, \bibinfo {author} {\bibfnamefont {C.~I.}\ \bibnamefont {Ip}}, \ and\ \bibinfo {author} {\bibfnamefont {S.~R.}\ \bibnamefont {Nagel}},\ }\href {\doibase 10.1103/PhysRevLett.130.197201} {\bibfield  {journal} {\bibinfo  {journal} {Physical Review Letters}\ }\textbf {\bibinfo {volume} {130}},\ \bibinfo {pages} {197201} (\bibinfo {year} {2023})}\BibitemShut {NoStop}%
\bibitem [{\citenamefont {Jules}\ \emph {et~al.}(2023)\citenamefont {Jules}, \citenamefont {Michel}, \citenamefont {Douin},\ and\ \citenamefont {Lechenault}}]{Jules2023}%
  \BibitemOpen
  \bibfield  {author} {\bibinfo {author} {\bibfnamefont {T.}~\bibnamefont {Jules}}, \bibinfo {author} {\bibfnamefont {L.}~\bibnamefont {Michel}}, \bibinfo {author} {\bibfnamefont {A.}~\bibnamefont {Douin}}, \ and\ \bibinfo {author} {\bibfnamefont {F.}~\bibnamefont {Lechenault}},\ }\href@noop {} {\bibfield  {journal} {\bibinfo  {journal} {Communications Physics}\ }\textbf {\bibinfo {volume} {6}} (\bibinfo {year} {2023})}\BibitemShut {NoStop}%
\bibitem [{\citenamefont {Gutierrez-Prieto}\ \emph {et~al.}(2025)\citenamefont {Gutierrez-Prieto}, \citenamefont {Meulblok}, \citenamefont {van Hecke},\ and\ \citenamefont {Reis}}]{Gutierrezprieto2025}%
  \BibitemOpen
  \bibfield  {author} {\bibinfo {author} {\bibfnamefont {E.}~\bibnamefont {Gutierrez-Prieto}}, \bibinfo {author} {\bibfnamefont {C.~M.}\ \bibnamefont {Meulblok}}, \bibinfo {author} {\bibfnamefont {M.}~\bibnamefont {van Hecke}}, \ and\ \bibinfo {author} {\bibfnamefont {P.~M.}\ \bibnamefont {Reis}},\ }\href {https://arxiv.org/abs/2508.16257} {\enquote {\bibinfo {title} {Dynamic driving enables independent control of material bits for targeted memory},}\ } (\bibinfo {year} {2025}),\ \Eprint {http://arxiv.org/abs/2508.16257} {arXiv:2508.16257 [cond-mat.soft]} \BibitemShut {NoStop}%
\bibitem [{\citenamefont {Brooks}\ \emph {et~al.}(1999)\citenamefont {Brooks}, \citenamefont {Fuller}, \citenamefont {Frank},\ and\ \citenamefont {Robertson}}]{Brooks1999}%
  \BibitemOpen
  \bibfield  {author} {\bibinfo {author} {\bibfnamefont {C.~F.}\ \bibnamefont {Brooks}}, \bibinfo {author} {\bibfnamefont {G.~G.}\ \bibnamefont {Fuller}}, \bibinfo {author} {\bibfnamefont {C.~W.}\ \bibnamefont {Frank}}, \ and\ \bibinfo {author} {\bibfnamefont {C.~R.}\ \bibnamefont {Robertson}},\ }\href {\doibase 10.1021/la980465r} {\bibfield  {journal} {\bibinfo  {journal} {Langmuir}\ }\textbf {\bibinfo {volume} {15}},\ \bibinfo {pages} {2450} (\bibinfo {year} {1999})}\BibitemShut {NoStop}%
\bibitem [{\citenamefont {Qiao}\ \emph {et~al.}(2021)\citenamefont {Qiao}, \citenamefont {Fan}, \citenamefont {Liu}, \citenamefont {Medina}, \citenamefont {Keim},\ and\ \citenamefont {Cheng}}]{Qiao2021}%
  \BibitemOpen
  \bibfield  {author} {\bibinfo {author} {\bibfnamefont {Y.}~\bibnamefont {Qiao}}, \bibinfo {author} {\bibfnamefont {C.}~\bibnamefont {Fan}}, \bibinfo {author} {\bibfnamefont {Z.}~\bibnamefont {Liu}}, \bibinfo {author} {\bibfnamefont {D.}~\bibnamefont {Medina}}, \bibinfo {author} {\bibfnamefont {N.~C.}\ \bibnamefont {Keim}}, \ and\ \bibinfo {author} {\bibfnamefont {X.}~\bibnamefont {Cheng}},\ }\href {\doibase 10.1122/8.0000263} {\bibfield  {journal} {\bibinfo  {journal} {Journal of Rheology}\ }\textbf {\bibinfo {volume} {65}},\ \bibinfo {pages} {1103} (\bibinfo {year} {2021})}\BibitemShut {NoStop}%
\bibitem [{\citenamefont {Keim}\ and\ \citenamefont {Medina}(2022)}]{Keim2022}%
  \BibitemOpen
  \bibfield  {author} {\bibinfo {author} {\bibfnamefont {N.~C.}\ \bibnamefont {Keim}}\ and\ \bibinfo {author} {\bibfnamefont {D.}~\bibnamefont {Medina}},\ }\href {\doibase 10.1126/sciadv.abo1614} {\bibfield  {journal} {\bibinfo  {journal} {Science Advances}\ }\textbf {\bibinfo {volume} {8}} (\bibinfo {year} {2022}),\ 10.1126/sciadv.abo1614}\BibitemShut {NoStop}%
\bibitem [{\citenamefont {Allan}\ \emph {et~al.}(2024)\citenamefont {Allan}, \citenamefont {Caswell}, \citenamefont {Keim}, \citenamefont {van~der Wel},\ and\ \citenamefont {Verweij}}]{trackpy}%
  \BibitemOpen
  \bibfield  {author} {\bibinfo {author} {\bibfnamefont {D.~B.}\ \bibnamefont {Allan}}, \bibinfo {author} {\bibfnamefont {T.}~\bibnamefont {Caswell}}, \bibinfo {author} {\bibfnamefont {N.~C.}\ \bibnamefont {Keim}}, \bibinfo {author} {\bibfnamefont {C.~M.}\ \bibnamefont {van~der Wel}}, \ and\ \bibinfo {author} {\bibfnamefont {R.~W.}\ \bibnamefont {Verweij}},\ }\href {\doibase 10.5281/zenodo.12708864} {\enquote {\bibinfo {title} {soft-matter/trackpy: v0.6.4},}\ } (\bibinfo {year} {2024})\BibitemShut {NoStop}%
\bibitem [{\citenamefont {Lundberg}\ \emph {et~al.}(2008)\citenamefont {Lundberg}, \citenamefont {Krishan}, \citenamefont {Xu}, \citenamefont {O’Hern},\ and\ \citenamefont {Dennin}}]{Lundberg2008}%
  \BibitemOpen
  \bibfield  {author} {\bibinfo {author} {\bibfnamefont {M.}~\bibnamefont {Lundberg}}, \bibinfo {author} {\bibfnamefont {K.}~\bibnamefont {Krishan}}, \bibinfo {author} {\bibfnamefont {N.}~\bibnamefont {Xu}}, \bibinfo {author} {\bibfnamefont {C.~S.}\ \bibnamefont {O’Hern}}, \ and\ \bibinfo {author} {\bibfnamefont {M.}~\bibnamefont {Dennin}},\ }\href {\doibase 10.1103/PhysRevE.77.041505} {\bibfield  {journal} {\bibinfo  {journal} {Physical Review E}\ }\textbf {\bibinfo {volume} {77}},\ \bibinfo {pages} {041505} (\bibinfo {year} {2008})}\BibitemShut {NoStop}%
\bibitem [{\citenamefont {Manning}\ and\ \citenamefont {Liu}(2011)}]{Manning2011}%
  \BibitemOpen
  \bibfield  {author} {\bibinfo {author} {\bibfnamefont {M.~L.}\ \bibnamefont {Manning}}\ and\ \bibinfo {author} {\bibfnamefont {A.~J.}\ \bibnamefont {Liu}},\ }\href {\doibase 10.1103/PhysRevLett.107.108302} {\bibfield  {journal} {\bibinfo  {journal} {Physical Review Letters}\ }\textbf {\bibinfo {volume} {107}},\ \bibinfo {pages} {108302} (\bibinfo {year} {2011})}\BibitemShut {NoStop}%
\bibitem [{\citenamefont {Richard}\ \emph {et~al.}(2020)\citenamefont {Richard}, \citenamefont {Ozawa}, \citenamefont {Patinet}, \citenamefont {Stanifer}, \citenamefont {Shang}, \citenamefont {Ridout}, \citenamefont {Xu}, \citenamefont {Zhang}, \citenamefont {Morse}, \citenamefont {Barrat}, \citenamefont {Berthier}, \citenamefont {Falk}, \citenamefont {Guan}, \citenamefont {Liu}, \citenamefont {Martens}, \citenamefont {Sastry}, \citenamefont {Vandembroucq}, \citenamefont {Lerner},\ and\ \citenamefont {Manning}}]{Richard2020}%
  \BibitemOpen
  \bibfield  {author} {\bibinfo {author} {\bibfnamefont {D.}~\bibnamefont {Richard}}, \bibinfo {author} {\bibfnamefont {M.}~\bibnamefont {Ozawa}}, \bibinfo {author} {\bibfnamefont {S.}~\bibnamefont {Patinet}}, \bibinfo {author} {\bibfnamefont {E.}~\bibnamefont {Stanifer}}, \bibinfo {author} {\bibfnamefont {B.}~\bibnamefont {Shang}}, \bibinfo {author} {\bibfnamefont {S.~A.}\ \bibnamefont {Ridout}}, \bibinfo {author} {\bibfnamefont {B.}~\bibnamefont {Xu}}, \bibinfo {author} {\bibfnamefont {G.}~\bibnamefont {Zhang}}, \bibinfo {author} {\bibfnamefont {P.~K.}\ \bibnamefont {Morse}}, \bibinfo {author} {\bibfnamefont {J.-L.}\ \bibnamefont {Barrat}}, \bibinfo {author} {\bibfnamefont {L.}~\bibnamefont {Berthier}}, \bibinfo {author} {\bibfnamefont {M.~L.}\ \bibnamefont {Falk}}, \bibinfo {author} {\bibfnamefont {P.}~\bibnamefont {Guan}}, \bibinfo {author} {\bibfnamefont {A.~J.}\ \bibnamefont {Liu}}, \bibinfo {author} {\bibfnamefont {K.}~\bibnamefont {Martens}}, \bibinfo {author} {\bibfnamefont {S.}~\bibnamefont {Sastry}}, \bibinfo {author} {\bibfnamefont {D.}~\bibnamefont {Vandembroucq}}, \bibinfo {author} {\bibfnamefont {E.}~\bibnamefont {Lerner}}, \ and\ \bibinfo {author} {\bibfnamefont {M.~L.}\ \bibnamefont {Manning}},\ }\href {\doibase 10.1103/PhysRevMaterials.4.113609} {\bibfield  {journal} {\bibinfo  {journal} {Physical Review Materials}\ }\textbf {\bibinfo {volume} {4}},\ \bibinfo {pages} {113609} (\bibinfo {year} {2020})}\BibitemShut {NoStop}%
\bibitem [{\citenamefont {Keim}\ and\ \citenamefont {Arratia}(2013)}]{Keim2013}%
  \BibitemOpen
  \bibfield  {author} {\bibinfo {author} {\bibfnamefont {N.~C.}\ \bibnamefont {Keim}}\ and\ \bibinfo {author} {\bibfnamefont {P.~E.}\ \bibnamefont {Arratia}},\ }\href {\doibase 10.1039/c3sm51014j} {\bibfield  {journal} {\bibinfo  {journal} {Soft Matter}\ }\textbf {\bibinfo {volume} {9}},\ \bibinfo {pages} {6222} (\bibinfo {year} {2013})}\BibitemShut {NoStop}%
\bibitem [{\citenamefont {Keim}\ \emph {et~al.}(2020)\citenamefont {Keim}, \citenamefont {Hass}, \citenamefont {Kroger},\ and\ \citenamefont {Wieker}}]{Keim2020}%
  \BibitemOpen
  \bibfield  {author} {\bibinfo {author} {\bibfnamefont {N.~C.}\ \bibnamefont {Keim}}, \bibinfo {author} {\bibfnamefont {J.}~\bibnamefont {Hass}}, \bibinfo {author} {\bibfnamefont {B.}~\bibnamefont {Kroger}}, \ and\ \bibinfo {author} {\bibfnamefont {D.}~\bibnamefont {Wieker}},\ }\href {\doibase 10.1103/PhysRevResearch.2.012004} {\bibfield  {journal} {\bibinfo  {journal} {Physical Review Research}\ }\textbf {\bibinfo {volume} {2}},\ \bibinfo {pages} {012004} (\bibinfo {year} {2020})}\BibitemShut {NoStop}%
\bibitem [{\citenamefont {Keim}(2014)}]{philatracks}%
  \BibitemOpen
  \bibfield  {author} {\bibinfo {author} {\bibfnamefont {N.~C.}\ \bibnamefont {Keim}},\ }\href {\doibase doi: 10.5281/zenodo.11459} {\enquote {\bibinfo {title} {philatracks v0.2},}\ } (\bibinfo {year} {2014})\BibitemShut {NoStop}%
\bibitem [{sup()}]{supp}%
  \BibitemOpen
  \href@noop {} {}\bibinfo {note} {See Supplemental Material at [URL will be inserted by publisher] for details on sample preparation, experimental protocols, data processing, model parameters, and a parameter sweep on the numerical model.}\BibitemShut {Stop}%
\bibitem [{\citenamefont {Galloway}\ \emph {et~al.}(2021)\citenamefont {Galloway}, \citenamefont {Teich}, \citenamefont {Ma}, \citenamefont {Kammer}, \citenamefont {Graham}, \citenamefont {Keim}, \citenamefont {Reina}, \citenamefont {Jerolmack}, \citenamefont {Yodh},\ and\ \citenamefont {Arratia}}]{Galloway2021}%
  \BibitemOpen
  \bibfield  {author} {\bibinfo {author} {\bibfnamefont {K.~L.}\ \bibnamefont {Galloway}}, \bibinfo {author} {\bibfnamefont {E.~G.}\ \bibnamefont {Teich}}, \bibinfo {author} {\bibfnamefont {X.~G.}\ \bibnamefont {Ma}}, \bibinfo {author} {\bibfnamefont {C.}~\bibnamefont {Kammer}}, \bibinfo {author} {\bibfnamefont {I.~R.}\ \bibnamefont {Graham}}, \bibinfo {author} {\bibfnamefont {N.~C.}\ \bibnamefont {Keim}}, \bibinfo {author} {\bibfnamefont {C.}~\bibnamefont {Reina}}, \bibinfo {author} {\bibfnamefont {D.~J.}\ \bibnamefont {Jerolmack}}, \bibinfo {author} {\bibfnamefont {A.~G.}\ \bibnamefont {Yodh}}, \ and\ \bibinfo {author} {\bibfnamefont {P.~E.}\ \bibnamefont {Arratia}},\ }\href {\doibase 10.1038/s41567-022-01536-9} {\bibfield  {journal} {\bibinfo  {journal} {Nature Physics}\ } (\bibinfo {year} {2021}),\ 10.1038/s41567-022-01536-9}\BibitemShut {NoStop}%
\bibitem [{\citenamefont {Adhikari}\ \emph {et~al.}(2022)\citenamefont {Adhikari}, \citenamefont {Mungan},\ and\ \citenamefont {Sastry}}]{Adhikari2022}%
  \BibitemOpen
  \bibfield  {author} {\bibinfo {author} {\bibfnamefont {M.}~\bibnamefont {Adhikari}}, \bibinfo {author} {\bibfnamefont {M.}~\bibnamefont {Mungan}}, \ and\ \bibinfo {author} {\bibfnamefont {S.}~\bibnamefont {Sastry}},\ }\href {https://arxiv.org/abs/2201.06535} {\enquote {\bibinfo {title} {Yielding behavior of glasses under asymmetric cyclic deformation},}\ } (\bibinfo {year} {2022}),\ \Eprint {http://arxiv.org/abs/2201.06535} {arXiv:2201.06535 [cond-mat.soft]} \BibitemShut {NoStop}%
\bibitem [{\citenamefont {Keim}\ and\ \citenamefont {Arratia}(2014)}]{Keim2014}%
  \BibitemOpen
  \bibfield  {author} {\bibinfo {author} {\bibfnamefont {N.~C.}\ \bibnamefont {Keim}}\ and\ \bibinfo {author} {\bibfnamefont {P.~E.}\ \bibnamefont {Arratia}},\ }\href {\doibase 10.1103/PhysRevLett.112.028302} {\bibfield  {journal} {\bibinfo  {journal} {Physical Review Letters}\ }\textbf {\bibinfo {volume} {112}},\ \bibinfo {pages} {028302} (\bibinfo {year} {2014})}\BibitemShut {NoStop}%
\bibitem [{\citenamefont {Keim}\ and\ \citenamefont {Arratia}(2015)}]{Keim2015}%
  \BibitemOpen
  \bibfield  {author} {\bibinfo {author} {\bibfnamefont {N.~C.}\ \bibnamefont {Keim}}\ and\ \bibinfo {author} {\bibfnamefont {P.~E.}\ \bibnamefont {Arratia}},\ }\href {\doibase 10.1039/C4SM02446J} {\bibfield  {journal} {\bibinfo  {journal} {Soft Matter}\ }\textbf {\bibinfo {volume} {11}},\ \bibinfo {pages} {1539} (\bibinfo {year} {2015})}\BibitemShut {NoStop}%
\bibitem [{\citenamefont {Rao}\ \emph {et~al.}(1990)\citenamefont {Rao}, \citenamefont {Krishnamurthy},\ and\ \citenamefont {Pandit}}]{Rao1990}%
  \BibitemOpen
  \bibfield  {author} {\bibinfo {author} {\bibfnamefont {M.}~\bibnamefont {Rao}}, \bibinfo {author} {\bibfnamefont {H.~R.}\ \bibnamefont {Krishnamurthy}}, \ and\ \bibinfo {author} {\bibfnamefont {R.}~\bibnamefont {Pandit}},\ }\href {\doibase 10.1103/PhysRevB.42.856} {\bibfield  {journal} {\bibinfo  {journal} {Physical Review B}\ }\textbf {\bibinfo {volume} {42}},\ \bibinfo {pages} {856} (\bibinfo {year} {1990})}\BibitemShut {NoStop}%
\bibitem [{\citenamefont {Jung}\ \emph {et~al.}(1990)\citenamefont {Jung}, \citenamefont {Gray}, \citenamefont {Roy},\ and\ \citenamefont {Mandel}}]{Jung1990}%
  \BibitemOpen
  \bibfield  {author} {\bibinfo {author} {\bibfnamefont {P.}~\bibnamefont {Jung}}, \bibinfo {author} {\bibfnamefont {G.}~\bibnamefont {Gray}}, \bibinfo {author} {\bibfnamefont {R.}~\bibnamefont {Roy}}, \ and\ \bibinfo {author} {\bibfnamefont {P.}~\bibnamefont {Mandel}},\ }\href {\doibase 10.1103/PhysRevLett.65.1873} {\bibfield  {journal} {\bibinfo  {journal} {Physical Review Letters}\ }\textbf {\bibinfo {volume} {65}},\ \bibinfo {pages} {1873} (\bibinfo {year} {1990})}\BibitemShut {NoStop}%
\bibitem [{\citenamefont {Preisach}(1935)}]{Preisach1935}%
  \BibitemOpen
  \bibfield  {author} {\bibinfo {author} {\bibfnamefont {F.}~\bibnamefont {Preisach}},\ }\href {\doibase 10.1007/BF01349418} {\bibfield  {journal} {\bibinfo  {journal} {Zeitschrift für Physik}\ }\textbf {\bibinfo {volume} {94}},\ \bibinfo {pages} {277} (\bibinfo {year} {1935})}\BibitemShut {NoStop}%
\bibitem [{\citenamefont {Keim}\ \emph {et~al.}(2019)\citenamefont {Keim}, \citenamefont {Paulsen}, \citenamefont {Zeravcic}, \citenamefont {Sastry},\ and\ \citenamefont {Nagel}}]{Keim2019}%
  \BibitemOpen
  \bibfield  {author} {\bibinfo {author} {\bibfnamefont {N.~C.}\ \bibnamefont {Keim}}, \bibinfo {author} {\bibfnamefont {J.~D.}\ \bibnamefont {Paulsen}}, \bibinfo {author} {\bibfnamefont {Z.}~\bibnamefont {Zeravcic}}, \bibinfo {author} {\bibfnamefont {S.}~\bibnamefont {Sastry}}, \ and\ \bibinfo {author} {\bibfnamefont {S.~R.}\ \bibnamefont {Nagel}},\ }\href {\doibase 10.1103/RevModPhys.91.035002} {\bibfield  {journal} {\bibinfo  {journal} {Reviews of Modern Physics}\ }\textbf {\bibinfo {volume} {91}},\ \bibinfo {pages} {035002} (\bibinfo {year} {2019})}\BibitemShut {NoStop}%
\bibitem [{\citenamefont {Mungan}\ \emph {et~al.}(2019)\citenamefont {Mungan}, \citenamefont {Sastry}, \citenamefont {Dahmen},\ and\ \citenamefont {Regev}}]{Mungan2019}%
  \BibitemOpen
  \bibfield  {author} {\bibinfo {author} {\bibfnamefont {M.}~\bibnamefont {Mungan}}, \bibinfo {author} {\bibfnamefont {S.}~\bibnamefont {Sastry}}, \bibinfo {author} {\bibfnamefont {K.}~\bibnamefont {Dahmen}}, \ and\ \bibinfo {author} {\bibfnamefont {I.}~\bibnamefont {Regev}},\ }\href@noop {} {\bibfield  {journal} {\bibinfo  {journal} {Phys. Rev. Lett.}\ }\textbf {\bibinfo {volume} {123}},\ \bibinfo {pages} {178002} (\bibinfo {year} {2019})}\BibitemShut {NoStop}%
\bibitem [{\citenamefont {Broner}\ \emph {et~al.}(1997)\citenamefont {Broner}, \citenamefont {Goldsztein},\ and\ \citenamefont {Strogatz}}]{Broner1997}%
  \BibitemOpen
  \bibfield  {author} {\bibinfo {author} {\bibfnamefont {F.}~\bibnamefont {Broner}}, \bibinfo {author} {\bibfnamefont {G.~H.}\ \bibnamefont {Goldsztein}}, \ and\ \bibinfo {author} {\bibfnamefont {S.~H.}\ \bibnamefont {Strogatz}},\ }\href {\doibase 10.1137/S0036139995290733} {\bibfield  {journal} {\bibinfo  {journal} {SIAM Journal on Applied Mathematics}\ }\textbf {\bibinfo {volume} {57}},\ \bibinfo {pages} {1163} (\bibinfo {year} {1997})}\BibitemShut {NoStop}%
\bibitem [{\citenamefont {Strogatz}(2018)}]{Strogatz2018}%
  \BibitemOpen
  \bibfield  {author} {\bibinfo {author} {\bibfnamefont {S.~H.}\ \bibnamefont {Strogatz}},\ }\href {\doibase 10.1201/9780429492563} {\emph {\bibinfo {title} {Nonlinear Dynamics and Chaos}}}\ (\bibinfo  {publisher} {CRC Press},\ \bibinfo {year} {2018})\BibitemShut {NoStop}%
\bibitem [{\citenamefont {Szulc}\ \emph {et~al.}(2022)\citenamefont {Szulc}, \citenamefont {Mungan},\ and\ \citenamefont {Regev}}]{Szulc2022}%
  \BibitemOpen
  \bibfield  {author} {\bibinfo {author} {\bibfnamefont {A.}~\bibnamefont {Szulc}}, \bibinfo {author} {\bibfnamefont {M.}~\bibnamefont {Mungan}}, \ and\ \bibinfo {author} {\bibfnamefont {I.}~\bibnamefont {Regev}},\ }\href {\doibase 10.1063/5.0087164} {\bibfield  {journal} {\bibinfo  {journal} {The Journal of Chemical Physics}\ }\textbf {\bibinfo {volume} {156}},\ \bibinfo {pages} {164506} (\bibinfo {year} {2022})}\BibitemShut {NoStop}%
\bibitem [{\citenamefont {Vasisht}\ \emph {et~al.}(2018)\citenamefont {Vasisht}, \citenamefont {Dutta}, \citenamefont {Gado},\ and\ \citenamefont {Blair}}]{Vasisht2018}%
  \BibitemOpen
  \bibfield  {author} {\bibinfo {author} {\bibfnamefont {V.~V.}\ \bibnamefont {Vasisht}}, \bibinfo {author} {\bibfnamefont {S.~K.}\ \bibnamefont {Dutta}}, \bibinfo {author} {\bibfnamefont {E.~D.}\ \bibnamefont {Gado}}, \ and\ \bibinfo {author} {\bibfnamefont {D.~L.}\ \bibnamefont {Blair}},\ }\href {\doibase 10.1103/PhysRevLett.120.018001} {\bibfield  {journal} {\bibinfo  {journal} {Physical Review Letters}\ }\textbf {\bibinfo {volume} {120}},\ \bibinfo {pages} {018001} (\bibinfo {year} {2018})}\BibitemShut {NoStop}%
\bibitem [{\citenamefont {Singh}\ \emph {et~al.}(2020)\citenamefont {Singh}, \citenamefont {Ozawa},\ and\ \citenamefont {Berthier}}]{Singh2020}%
  \BibitemOpen
  \bibfield  {author} {\bibinfo {author} {\bibfnamefont {M.}~\bibnamefont {Singh}}, \bibinfo {author} {\bibfnamefont {M.}~\bibnamefont {Ozawa}}, \ and\ \bibinfo {author} {\bibfnamefont {L.}~\bibnamefont {Berthier}},\ }\href {\doibase 10.1103/PhysRevMaterials.4.025603} {\bibfield  {journal} {\bibinfo  {journal} {Physical Review Materials}\ }\textbf {\bibinfo {volume} {4}},\ \bibinfo {pages} {025603} (\bibinfo {year} {2020})}\BibitemShut {NoStop}%
\bibitem [{\citenamefont {Lerner}\ and\ \citenamefont {Bouchbinder}(2021)}]{Lerner2021}%
  \BibitemOpen
  \bibfield  {author} {\bibinfo {author} {\bibfnamefont {E.}~\bibnamefont {Lerner}}\ and\ \bibinfo {author} {\bibfnamefont {E.}~\bibnamefont {Bouchbinder}},\ }\href {\doibase 10.1063/5.0069477} {\bibfield  {journal} {\bibinfo  {journal} {The Journal of Chemical Physics}\ }\textbf {\bibinfo {volume} {155}} (\bibinfo {year} {2021}),\ 10.1063/5.0069477}\BibitemShut {NoStop}%
\bibitem [{\citenamefont {Amir}\ \emph {et~al.}(2012)\citenamefont {Amir}, \citenamefont {Oreg},\ and\ \citenamefont {Imry}}]{Amir2012}%
  \BibitemOpen
  \bibfield  {author} {\bibinfo {author} {\bibfnamefont {A.}~\bibnamefont {Amir}}, \bibinfo {author} {\bibfnamefont {Y.}~\bibnamefont {Oreg}}, \ and\ \bibinfo {author} {\bibfnamefont {Y.}~\bibnamefont {Imry}},\ }\href {\doibase 10.1073/pnas.1120147109} {\bibfield  {journal} {\bibinfo  {journal} {Proceedings of the National Academy of Sciences}\ }\textbf {\bibinfo {volume} {109}},\ \bibinfo {pages} {1850} (\bibinfo {year} {2012})}\BibitemShut {NoStop}%
\bibitem [{\citenamefont {Szulc}\ and\ \citenamefont {Regev}(2024)}]{Szulc2024}%
  \BibitemOpen
  \bibfield  {author} {\bibinfo {author} {\bibfnamefont {A.}~\bibnamefont {Szulc}}\ and\ \bibinfo {author} {\bibfnamefont {I.}~\bibnamefont {Regev}},\ }\href {\doibase 10.1103/PhysRevResearch.6.033129} {\bibfield  {journal} {\bibinfo  {journal} {Physical Review Research}\ }\textbf {\bibinfo {volume} {6}},\ \bibinfo {pages} {033129} (\bibinfo {year} {2024})}\BibitemShut {NoStop}%
\bibitem [{\citenamefont {Jagla}(2007)}]{Jagla2007}%
  \BibitemOpen
  \bibfield  {author} {\bibinfo {author} {\bibfnamefont {E.~A.}\ \bibnamefont {Jagla}},\ }\href {\doibase 10.1103/PhysRevE.76.046119} {\bibfield  {journal} {\bibinfo  {journal} {Physical Review E}\ }\textbf {\bibinfo {volume} {76}},\ \bibinfo {pages} {046119} (\bibinfo {year} {2007})}\BibitemShut {NoStop}%
\bibitem [{\citenamefont {Martens}\ \emph {et~al.}(2012)\citenamefont {Martens}, \citenamefont {Bocquet},\ and\ \citenamefont {Barrat}}]{Martens2012}%
  \BibitemOpen
  \bibfield  {author} {\bibinfo {author} {\bibfnamefont {K.}~\bibnamefont {Martens}}, \bibinfo {author} {\bibfnamefont {L.}~\bibnamefont {Bocquet}}, \ and\ \bibinfo {author} {\bibfnamefont {J.-L.}\ \bibnamefont {Barrat}},\ }\href {\doibase 10.1039/c2sm07090a} {\bibfield  {journal} {\bibinfo  {journal} {Soft Matter}\ }\textbf {\bibinfo {volume} {8}},\ \bibinfo {pages} {4197} (\bibinfo {year} {2012})}\BibitemShut {NoStop}%
\bibitem [{\citenamefont {Cubuk}\ \emph {et~al.}(2017)\citenamefont {Cubuk}, \citenamefont {Ivancic}, \citenamefont {Schoenholz}, \citenamefont {Strickland}, \citenamefont {Basu}, \citenamefont {Davidson}, \citenamefont {Fontaine}, \citenamefont {Hor}, \citenamefont {Huang}, \citenamefont {Jiang}, \citenamefont {Keim}, \citenamefont {Koshigan}, \citenamefont {Lefever}, \citenamefont {Liu}, \citenamefont {Ma}, \citenamefont {Magagnosc}, \citenamefont {Morrow}, \citenamefont {Ortiz}, \citenamefont {Rieser}, \citenamefont {Shavit}, \citenamefont {Still}, \citenamefont {Xu}, \citenamefont {Zhang}, \citenamefont {Nordstrom}, \citenamefont {Arratia}, \citenamefont {Carpick}, \citenamefont {Durian}, \citenamefont {Fakhraai}, \citenamefont {Jerolmack}, \citenamefont {Lee}, \citenamefont {Li}, \citenamefont {Riggleman}, \citenamefont {Turner}, \citenamefont {Yodh}, \citenamefont {Gianola},\ and\ \citenamefont {Liu}}]{Cubuk2017}%
  \BibitemOpen
  \bibfield  {author} {\bibinfo {author} {\bibfnamefont {E.~D.}\ \bibnamefont {Cubuk}}, \bibinfo {author} {\bibfnamefont {R.~J.~S.}\ \bibnamefont {Ivancic}}, \bibinfo {author} {\bibfnamefont {S.~S.}\ \bibnamefont {Schoenholz}}, \bibinfo {author} {\bibfnamefont {D.~J.}\ \bibnamefont {Strickland}}, \bibinfo {author} {\bibfnamefont {A.}~\bibnamefont {Basu}}, \bibinfo {author} {\bibfnamefont {Z.~S.}\ \bibnamefont {Davidson}}, \bibinfo {author} {\bibfnamefont {J.}~\bibnamefont {Fontaine}}, \bibinfo {author} {\bibfnamefont {J.~L.}\ \bibnamefont {Hor}}, \bibinfo {author} {\bibfnamefont {Y.-R.}\ \bibnamefont {Huang}}, \bibinfo {author} {\bibfnamefont {Y.}~\bibnamefont {Jiang}}, \bibinfo {author} {\bibfnamefont {N.~C.}\ \bibnamefont {Keim}}, \bibinfo {author} {\bibfnamefont {K.~D.}\ \bibnamefont {Koshigan}}, \bibinfo {author} {\bibfnamefont {J.~A.}\ \bibnamefont {Lefever}}, \bibinfo {author} {\bibfnamefont {T.}~\bibnamefont {Liu}}, \bibinfo {author} {\bibfnamefont {X.-G.}\ \bibnamefont {Ma}}, \bibinfo {author} {\bibfnamefont {D.~J.}\ \bibnamefont {Magagnosc}}, \bibinfo {author} {\bibfnamefont {E.}~\bibnamefont {Morrow}}, \bibinfo {author} {\bibfnamefont {C.~P.}\ \bibnamefont {Ortiz}}, \bibinfo {author} {\bibfnamefont {J.~M.}\ \bibnamefont {Rieser}}, \bibinfo {author} {\bibfnamefont {A.}~\bibnamefont {Shavit}}, \bibinfo {author} {\bibfnamefont {T.}~\bibnamefont {Still}}, \bibinfo {author} {\bibfnamefont {Y.}~\bibnamefont {Xu}}, \bibinfo {author} {\bibfnamefont {Y.}~\bibnamefont {Zhang}}, \bibinfo {author} {\bibfnamefont {K.~N.}\ \bibnamefont {Nordstrom}}, \bibinfo {author} {\bibfnamefont {P.~E.}\ \bibnamefont {Arratia}}, \bibinfo {author} {\bibfnamefont {R.~W.}\ \bibnamefont {Carpick}}, \bibinfo {author} {\bibfnamefont {D.~J.}\ \bibnamefont {Durian}}, \bibinfo {author} {\bibfnamefont {Z.}~\bibnamefont {Fakhraai}}, \bibinfo {author} {\bibfnamefont {D.~J.}\ \bibnamefont {Jerolmack}}, \bibinfo {author} {\bibfnamefont {D.}~\bibnamefont {Lee}}, \bibinfo {author} {\bibfnamefont {J.}~\bibnamefont {Li}}, \bibinfo {author} {\bibfnamefont {R.}~\bibnamefont {Riggleman}}, \bibinfo {author} {\bibfnamefont {K.~T.}\ \bibnamefont {Turner}}, \bibinfo {author} {\bibfnamefont {A.~G.}\ \bibnamefont {Yodh}}, \bibinfo {author} {\bibfnamefont {D.~S.}\ \bibnamefont {Gianola}}, \ and\ \bibinfo {author} {\bibfnamefont {A.~J.}\ \bibnamefont {Liu}},\ }\href {\doibase 10.1126/science.aai8830} {\bibfield  {journal} {\bibinfo  {journal} {Science}\ }\textbf {\bibinfo {volume} {358}},\ \bibinfo {pages} {1033} (\bibinfo {year} {2017})}\BibitemShut {NoStop}%
\bibitem [{\citenamefont {Danov}\ and\ \citenamefont {Kralchevsky}(2010)}]{Danov2010}%
  \BibitemOpen
  \bibfield  {author} {\bibinfo {author} {\bibfnamefont {K.~D.}\ \bibnamefont {Danov}}\ and\ \bibinfo {author} {\bibfnamefont {P.~A.}\ \bibnamefont {Kralchevsky}},\ }\href {\doibase 10.1016/j.jcis.2010.02.017} {\bibfield  {journal} {\bibinfo  {journal} {Journal of Colloid and Interface Science}\ }\textbf {\bibinfo {volume} {345}},\ \bibinfo {pages} {505} (\bibinfo {year} {2010})}\BibitemShut {NoStop}%
\bibitem [{\citenamefont {Masschaele}\ \emph {et~al.}(2010)\citenamefont {Masschaele}, \citenamefont {Park}, \citenamefont {Furst}, \citenamefont {Fransaer},\ and\ \citenamefont {Vermant}}]{Masschaele2010}%
  \BibitemOpen
  \bibfield  {author} {\bibinfo {author} {\bibfnamefont {K.}~\bibnamefont {Masschaele}}, \bibinfo {author} {\bibfnamefont {B.~J.}\ \bibnamefont {Park}}, \bibinfo {author} {\bibfnamefont {E.~M.}\ \bibnamefont {Furst}}, \bibinfo {author} {\bibfnamefont {J.}~\bibnamefont {Fransaer}}, \ and\ \bibinfo {author} {\bibfnamefont {J.}~\bibnamefont {Vermant}},\ }\href {\doibase 10.1103/PhysRevLett.105.048303} {\bibfield  {journal} {\bibinfo  {journal} {Physical Review Letters}\ }\textbf {\bibinfo {volume} {105}},\ \bibinfo {pages} {048303} (\bibinfo {year} {2010})}\BibitemShut {NoStop}%
\bibitem [{\citenamefont {Chen}\ \emph {et~al.}(2005)\citenamefont {Chen}, \citenamefont {Tan}, \citenamefont {Ng}, \citenamefont {Ford},\ and\ \citenamefont {Tong}}]{Chen2005}%
  \BibitemOpen
  \bibfield  {author} {\bibinfo {author} {\bibfnamefont {W.}~\bibnamefont {Chen}}, \bibinfo {author} {\bibfnamefont {S.}~\bibnamefont {Tan}}, \bibinfo {author} {\bibfnamefont {T.-K.}\ \bibnamefont {Ng}}, \bibinfo {author} {\bibfnamefont {W.~T.}\ \bibnamefont {Ford}}, \ and\ \bibinfo {author} {\bibfnamefont {P.}~\bibnamefont {Tong}},\ }\href {\doibase 10.1103/PhysRevLett.95.218301} {\bibfield  {journal} {\bibinfo  {journal} {Physical Review Letters}\ }\textbf {\bibinfo {volume} {95}},\ \bibinfo {pages} {218301} (\bibinfo {year} {2005})}\BibitemShut {NoStop}%
\bibitem [{\citenamefont {Tajuelo}\ \emph {et~al.}(2016)\citenamefont {Tajuelo}, \citenamefont {Pastor},\ and\ \citenamefont {Rubio}}]{Tajuelo2016}%
  \BibitemOpen
  \bibfield  {author} {\bibinfo {author} {\bibfnamefont {J.}~\bibnamefont {Tajuelo}}, \bibinfo {author} {\bibfnamefont {J.~M.}\ \bibnamefont {Pastor}}, \ and\ \bibinfo {author} {\bibfnamefont {M.~A.}\ \bibnamefont {Rubio}},\ }\href {\doibase 10.1122/1.4958668} {\bibfield  {journal} {\bibinfo  {journal} {Journal of Rheology}\ }\textbf {\bibinfo {volume} {60}},\ \bibinfo {pages} {1095} (\bibinfo {year} {2016})}\BibitemShut {NoStop}%
\bibitem [{\citenamefont {Reynaert}\ \emph {et~al.}(2008)\citenamefont {Reynaert}, \citenamefont {Brooks}, \citenamefont {Moldenaers}, \citenamefont {Vermant},\ and\ \citenamefont {Fuller}}]{Reynaert2008}%
  \BibitemOpen
  \bibfield  {author} {\bibinfo {author} {\bibfnamefont {S.}~\bibnamefont {Reynaert}}, \bibinfo {author} {\bibfnamefont {C.~F.}\ \bibnamefont {Brooks}}, \bibinfo {author} {\bibfnamefont {P.}~\bibnamefont {Moldenaers}}, \bibinfo {author} {\bibfnamefont {J.}~\bibnamefont {Vermant}}, \ and\ \bibinfo {author} {\bibfnamefont {G.~G.}\ \bibnamefont {Fuller}},\ }\href@noop {} {\bibfield  {journal} {\bibinfo  {journal} {Journal of Rheology}\ }\textbf {\bibinfo {volume} {52}},\ \bibinfo {pages} {261} (\bibinfo {year} {2008})}\BibitemShut {NoStop}%
\end{thebibliography}%

\end{document}